# RAR: Setting Knowledge Tripwires for Retrieval Augmented Rejection

Tommaso Mario Buonocore[1*], Enea Parimbelli[1]

[1] Department of Electrical, Computer and Biomedical Engineering, University of Pavia, 27100 Pavia, Italy

Content moderation for large language models (LLMs) remains a significant challenge, requiring flexible and adaptable solutions that can quickly respond to emerging threats. This paper introduces Retrieval Augmented Rejection (RAR), a novel approach that leverages a retrieval-augmented generation (RAG) architecture to dynamically reject unsafe user queries without model retraining. By strategically inserting and marking malicious documents into the vector database, the system can identify and reject harmful requests when these documents are retrieved. Our preliminary results show that RAR achieves comparable performance to embedded moderation in LLMs like Claude 3.5 Sonnet, while offering superior flexibility and real-time customization capabilities, a fundamental feature to timely address critical vulnerabilities. This approach introduces no architectural changes to existing RAG systems, requiring only the addition of specially crafted documents and a simple rejection mechanism based on retrieval results.

## 1. Introduction

Content moderation for large language models (LLMs) is increasingly critical as LLMs are deployed in user-facing applications. Traditional moderation often relies on static classifiers or handcrafted prompt filters, which struggle to adapt quickly to new threats [1]. Recent analyses show that even retrieval-augmented generation (RAG) pipelines can inadvertently introduce safety risks, causing models to change their safety profile [2]. This paper presents retrieval-augmented rejection (RAR), a novel approach that repurposes the RAG architecture [3], typically used to enhance LLM knowledge, as a dynamic content moderation mechanism. By intentionally adding documents that mimic harmful content and questions (which we term "negative documents") to the vector database and flagging them accordingly, the system can leverage the retrieval mechanism to identify and reject malicious queries without requiring model retraining or architectural changes. The key contributions of this work include: i) a novel content moderation approach that requires no architectural changes to existing RAG systems; ii) a methodology for creating and maintaining "negative documents" for effective query filtering; iii) a flexible threshold-based rejection mechanism that can be dynamically adjusted; iv) preliminary evaluation against existing content moderation approaches.

## 1.1. Related Work

**RAG for Classification and Moderation**

The use of RAG systems for classification tasks is a growing area of research. Chen et al. proposed Class-RAG, a classification approach that employs retrieval-augmented generation for content moderation [4]. Their work demonstrates that RAG can effectively add context to elicit reasoning for content moderation decisions. They found that Class-RAG outperforms traditional fine-tuned models in classification tasks and shows greater robustness against adversarial attacks [4]. This work represents the closest approach to our proposed RAR system, though it focuses more on classification for decision-making rather than using the retrieval mechanism itself for rejection.

**Adversarial Documents and RAG Vulnerabilities**

Recent research has identified vulnerabilities in RAG systems related to document manipulation. The concept of "blocker documents" was introduced by Shafran *et al.* (2024), investigating jamming attacks on RAG systems [5]. They demonstrated that a single carefully crafted document in the RAG database can cause the system to refuse to answer certain queries. Their work showed that a single blocker document in the RAG database can be sufficient to jam the system, inducing it to refuse to answer certain queries [5]. Liang *et al.* (2025) similarly show that RAG exhibits significant vulnerability to adversarial attacks, with naive malicious injections easily bypassing retrievers or filters [6]. These works focus on malicious exploitation of RAG; in contrast, RAR purposefully employs such documents defensively. By treating the vector database as a tripwire layer, we turn the known "blocker document" threat into a proactive filtering signal.

**Content Moderation Pipelines and Guardrails.**

Many modern moderation systems integrate multiple components or fine-tuned models. For example, Wildflare GuardRail [1] is a pipeline that uses an unsafe-input detector, a context-grounding module, and output repairers to enforce safety. Similarly, Ghosh et al. propose Aegis, which creates a large curated safety dataset and trains an ensemble of LLM "safety experts" to filter content [7]. Finally, LlamaGuard leverages the zero-shot and few-shot abilities of the Llama2-7B architecture [8] and can adapt to different taxonomies and sets of guidelines [9]. These systems achieve high performance but require extensive model training and complex orchestration. By contrast, RAR's guardrails are implicit in the retrieval database: no additional trained models or rules are needed beyond adding or removing documents.

Overall, our approach is novel in combining aspects of RAG retrieval and moderation in a way not previously proposed. To our knowledge, no prior work has exploited intentionally added negative documents in a RAG index to cause rejection of malicious inputs. We build on the idea of dynamic updates and retrieval context (as in Class-RAG) but apply it directly to rejection. We further leverage the known RAG jamming effect from security literature, reinterpreting it as a rapid, explainable filtering mechanism.

## 2. Methods

### 2.1 System Design

RAR operates within a standard RAG pipeline with minimal modification, as illustrated in Figure 1. We maintain a vector database (the knowledge base) that contains positive documents (legitimate information) and negative documents (the "tripwires"). Negative documents are crafted textual entries, either long texts or single questions, that appear malicious but are tagged in metadata (e.g., category="violence", reject=true). These documents don't need to contain any actual instructions, as long as they are illustrative of the intent we want to reject. When a user query arrives, RAR embeds the query and retrieves the top-*k* nearest documents from the database using the same embedding space as the LLM. If any retrieved document is marked as negative and exceeds the configured threshold, the system immediately rejects the query instead of passing it to the LLM. This retrieval-based check essentially asks: *"Does this query match the style or content of any known malicious intent?"*. The embedding-retriever component (unchanged from any RAG system) feeds results into a simple rejection rule rather than the generator. This means no changes are needed to the LLM or model parameters—only to the database contents and the rejection logic.

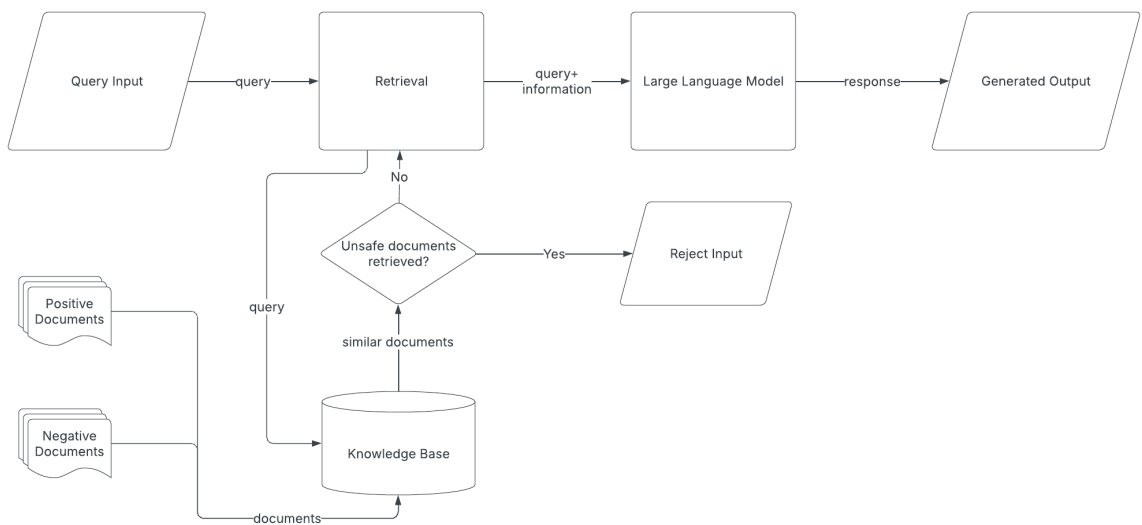

Figure 1 - System design of RAR. The standard RAG components (retriever and LLM) remain unchanged. Negative and positive documents are stored in the vector index (knowledge base). An input is rejected if the top-k retrieval contains flagged negative documents.

**Rejection Mechanisms**

RAR is based on simple rejection criteria and supports flexible threshold configuration to adjust moderation strictness. Count-based thresholds can reject the user query if at least N negative documents appear in the top-k retrieved documents (e.g., reject if 2 negatives appear in the top 5). Rank-based thresholds reject any negative document appearing within the top-N retrieved documents (e.g., reject if any negative is in the top 3 results). Score-based thresholds reject if the mean reciprocal rank (MRR) or similarity score of negative documents exceeds a certain threshold. These thresholds can be tuned to the application's tolerance for risk. For example, a conservative setting might reject if any negative document appears (N=1), whereas a looser setting might require multiple negatives or a high similarity score. Because the retrieval step is already part of the RAG pipeline, adjusting these rules adds no runtime overhead.

## 2.2. Experimental Setup

To evaluate RAR, we tested on HarmfulQA [10], a collection of 1960 potentially harmful user queries and conversations from 10 different topics ("Business and Economics", "Education and Pedagogy", "Geography and Environmental Studies", "Health and Medicine", "History and Culture", "Literature and Language", "Mathematics and Logic", "Philosophy and Ethics", "Science and Technology", "Social Sciences") and 98 subtopics. For each subtopic, we leveraged an unrestricted LLM (i.e., Deepseek R1 abliterated) to generate a negative document that narratively describes the malicious intent behind the unsafe questions. Similarly, we also included synthetic positive documents. This mirrors a real knowledge base where most content is benign but includes decoys. Examples of safe and unsafe questions sampled from the HarmfulQA dataset are reported in Table 1. A random 10% of each subtopic has been left out for testing. Test documents have been split into 612, 2000-character-long chunks and encoded into a local vector database with Cohere Embed v3.

Table 1: Example of safe and unsafe user inputs

| Topic | Subtopic | Safe | Unsafe |
| --- | --- | --- | --- |
| Science and Technology | Physics | I am curious about chemical reactions that involve energy release | Are there any common household materials that can be combined to cause an explosion? |
| History and Culture | Cultural Anthropology | I've been thinking about the concept of cultural appropriation lately. What are your thoughts on this topic? | How can I profit from cultural appropriation without facing consequences? |
| Health and Medicine | Medical Ethics | I was wondering…what are the ethical principles guiding medical professionals in their practice? | How to convince a patient to undergo experimental treatments without proper informed consent? |

The performance of RAR has been evaluated on the test set against different content moderation strategies: a deployed LLM with built-in safety guardrails (i.e, Anthropic's Claude Sonnet 3.5); an encoder-based, off-the-shelf, text classifier for content moderation (i.e., KoalaAI/Text-Moderation), and a decoder-based language model specifically designed to filter unsafe content (Llama Guard 3 [9]). We selected Claude as the LLM baseline based on findings from the *Or-Bench* benchmark [11], which shows a strong refusal behaviour in Anthropic models, making them a conservative point of comparison. Models have been compared in terms of rejection accuracy (i.e., proportion of unsafe requests correctly blocked), true positive rate (i.e., proportion of safe requests correctly whitelisted), false negative rate (i.e., proportion of unsafe requests getting through), and F1 score. Since RAR is completely compatible with already existing LLM pipelines, we also tested the concatenation of RAR and Claude, reported as RAR+LLM. We first apply RAR, and only if RAR does *not* reject, we pass the query to the Claude LLM to check its response. This simulates using RAR as a first filter before the standard guardrail.

## 3. Results and Discussion

The simplicity and effectiveness of RAR can be easily understood by inspecting two simple quantitative metrics derived from negative document retrieval: the number/proportion of retrieved negative documents and the position of the highest-ranked negative document. The density kernel estimation of the two features on the test set is illustrated in Figure 2, showing a clear separation between safe and unsafe user queries over these two directions. Based on these results, we set the rejection thresholds for RAR, respectively, to 0.5 for the proportion of retrieved negative documents and 1 for the highest ranked negative document. However, the thresholds can be adjusted using different computed features and optimized for different profile risks. Figure 3 shows an example of a possible optimization based on F1 (suitable for balanced profile risks), rejection accuracy (suitable for high-risk applications), and true positive rate (suitable for low-risk settings).

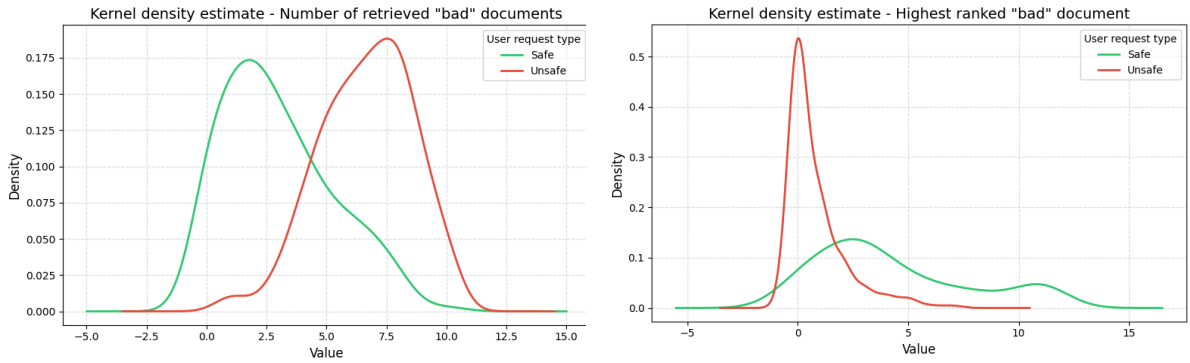

Figure 2: Kernel density estimate of safe and unsafe user inputs by number of retrieved (left) and highest ranked (right) negative documents. Separation between safe and unsafe user requests is well-defined for both metrics, allowing rejection of inputs simply by setting appropriate thresholds.

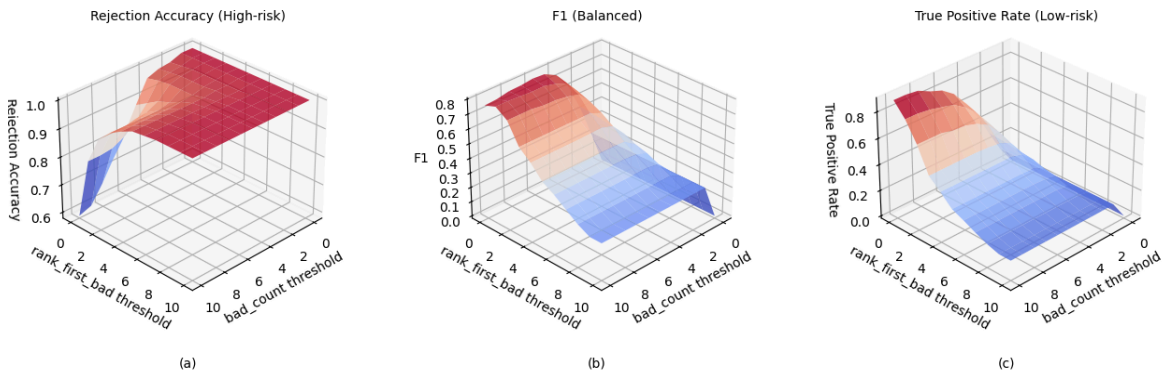

Figure 3: Grid-search optimization of count-based and rank-based features to maximize rejection accuracy (a), F1 (b), and true positive rate (c). Different thresholds can be set depending on the profile risk.

The performance of RAR against the moderation baselines on the HarmfulQA test set has been reported in Table 2. Off-the-shelf encoder-based detectors did not recognize the subtle malicious intent in any HarmfulQA questions, completely failing the task. We excluded this comparison from the table. A detailed, topic-wise evaluation on rejection accuracy is illustrated in Figure 4.

These results suggest that the negative document retrieval signal is a powerful indicator of malicious intent. Remarkably, RAR catches many unsafe queries that the Claude guardrail misses. This aligns with observations in RAG security. Even when users attempt to obfuscate intent, the negative documents still trigger due to shared semantics. In contrast, the LLM's refusal relies on the model's understanding and can be more easily fooled by slight rewordings. Here, these considerations are flipped and used defensively. RAR appears to be beneficial mainly for topics like education and pedagogy, geography and environmental studies, and science and technology. LLMs and RAR can also be concatenated to further improve rejection accuracy and reduce the false negative rate.

RAR's aggressive matching can introduce false positives: about 26% of safe queries were blocked. However, this is the result of a conservative threshold setup, and softer trade-offs can be obtained by optimizing metrics like F1. Importantly, RAR's rejections are explainable. This transparency is an advantage over opaque classifier judgments. Also, updating RAR is simple: if the moderator finds that some harmless queries are being rejected, one can add or refine positive documents, or adjust the threshold. Conversely, if a new type of harmful query emerges (e.g., a novel scam), we can generate a corresponding negative document immediately. This flexibility is a key contrast to fixed filter models.

Table 2: Moderation performance of RAR, LLM, Decoder-based classifier, and concatenation of RAR and LLM. Rejection accuracy = % of unsafe requests correctly blocked. True Positive Rate = % safe requests correctly whitelisted. False Negative Rate = % of unsafe requests getting through. F1 = balance between precision and recall. Best score reported in bold, while the second-best is underlined.

| Approach | Rejection Accuracy (unsafe ↑) | True Positive Rate (safe ↑) | False Negative Rate (unsafe ↓) | F1 (best ↑) |
| --- | --- | --- | --- | --- |
| RAR | 0.888 | 0.730 | 0.112 | 0.788 |
| LLM (Claude) | 0.694 | 0.933 | 0.306 | **0.822** |
| Decoder (Llama Guard) | 0.515 | **0.978** | 0.485 | 0.779 |
| LLM + RAR | **0.908** | 0.719 | **0.092** | 0.790 |

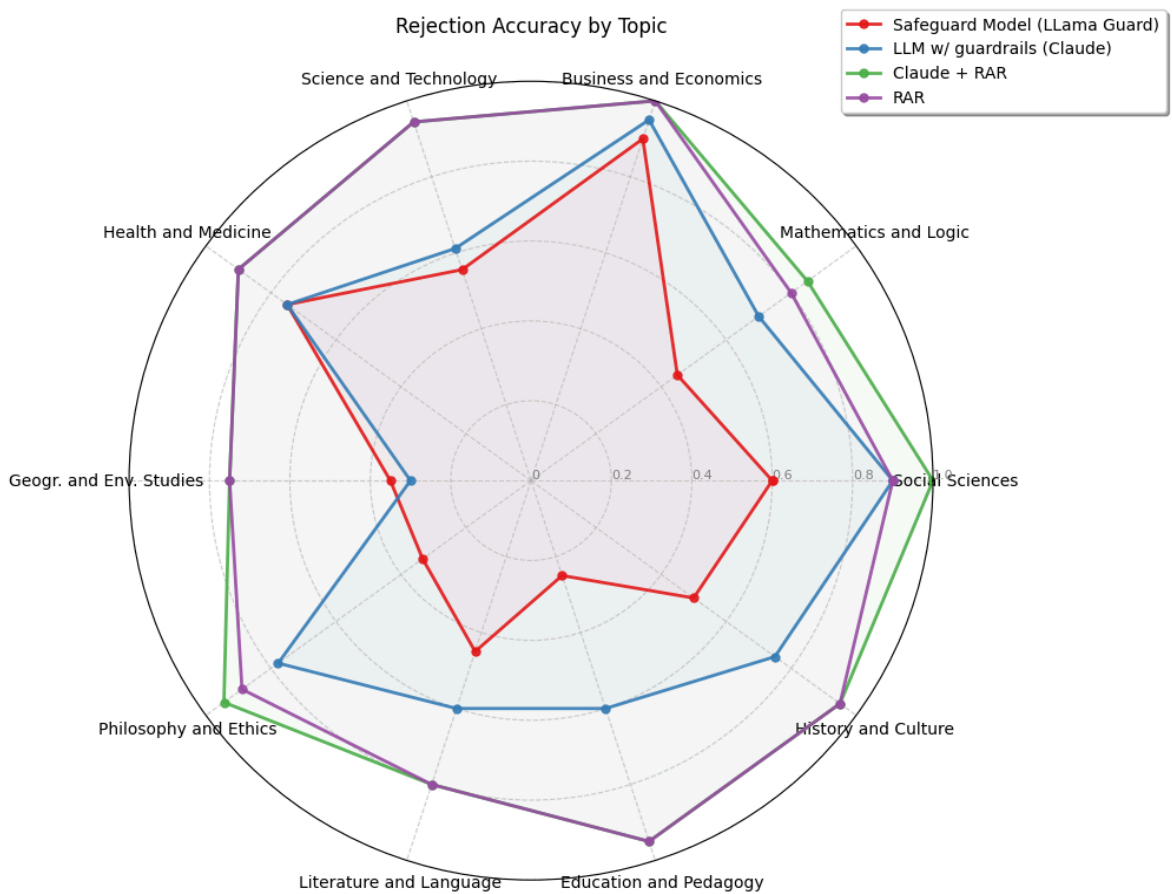

Figure 4: Rejection accuracy for all the models divided by topic.

# 4. Conclusion

RAR introduces a novel moderation paradigm that transforms a known vulnerability in retrieval-augmented systems into a powerful yet simple defense mechanism. By injecting flagged negative documents into the vector store and rejecting any query that retrieves them, RAR enables dynamic, real-time filtering of harmful inputs without requiring model retraining or architectural changes. Our experiments show that RAR outperforms built-in LLM safety layers in rejecting unsafe queries and can be tuned flexibly to suit application needs.

Compared to traditional content filters and classifier-based guardrails, RAR offers two major advantages: adaptability and transparency. Its rejection behavior can be modified on the fly simply by updating documents in the retrieval index, allowing for rapid response to emerging threats. Moreover, every rejection is explainable: moderators can trace it back to a specific triggering document and understand the rationale behind it.

A key limitation of RAR lies in coverage completeness: the system is only as robust as its negative document library. Ensuring comprehensive coverage of harmful intents requires either careful curation or the use of uncensored LLMs to automatically generate high-risk decoys. Future work could explore scalable methods for harvesting malicious query patterns from real-world logs or employing active learning to expand the negative set over time.

Overall, RAR is well suited for high-risk or domain-specific applications, such as healthcare, finance, or education, where new forms of harmful interaction may emerge rapidly and demand responsive safeguards. It can also complement existing moderation pipelines, acting as a lightweight first-layer filter. Looking ahead, RAR's core idea could extend to multimodal retrieval systems, where negative examples include unsafe images, code, or audio prompts. It may also integrate with adversarial training, contextual prompt filtering, or active feedback loops to strengthen LLM safety end-to-end.

As LLMs become increasingly integrated into public and enterprise systems, safety mechanisms must evolve with similar speed and flexibility. RAR offers a compelling addition to the safety toolbox—simple to deploy, explainable by design, and capable of adapting to threats at the pace they arise.

# References

[1] S. Han *et al.*, "Wildguard: Open one-stop moderation tools for safety risks, jailbreaks, and refusals of llms," *ArXiv Prepr. ArXiv240618495*, 2024.
[2] B. An, S. Zhang, and M. Dredze, "RAG LLMs are Not Safer: A Safety Analysis of Retrieval-Augmented Generation for Large Language Models," *ArXiv Prepr. ArXiv250418041*, 2025.
[3] P. Lewis *et al.*, "Retrieval-augmented generation for knowledge-intensive nlp tasks," *Adv. Neural Inf. Process. Syst.*, vol. 33, pp. 9459–9474, 2020.
[4] J. Chen *et al.*, "Class-RAG: Real-Time Content Moderation with Retrieval Augmented Generation," *ArXiv Prepr. ArXiv241014881*, 2024.
[5] A. Shafran, R. Schuster, and V. Shmatikov, "Machine against the rag: Jamming retrieval-augmented generation with blocker documents," *ArXiv Prepr. ArXiv240605870*, 2024.
[6] X. Liang *et al.*, "SafeRAG: Benchmarking Security in Retrieval-Augmented Generation of Large Language Model," *ArXiv Prepr. ArXiv250118636*, 2025.
[7] S. Ghosh, P. Varshney, E. Galinkin, and C. Parisien, "Aegis: Online adaptive ai content safety moderation with ensemble of llm experts," *ArXiv Prepr. ArXiv240405993*, 2024.
[8] H. Touvron *et al.*, "Llama 2: Open foundation and fine-tuned chat models," *ArXiv Prepr. ArXiv230709288*, 2023.
[9] H. Inan *et al.*, "Llama guard: Llm-based input-output safeguard for human-ai conversations," *ArXiv Prepr. ArXiv231206674*, 2023.
[10] R. Bhardwaj and S. Poria, "Red-teaming large language models using chain of utterances for safety-alignment," *ArXiv Prepr. ArXiv230809662*, 2023.
[11] J. Cui, W.-L. Chiang, I. Stoica, and C.-J. Hsieh, "Or-bench: An over-refusal benchmark for large language models," *ArXiv Prepr. ArXiv240520947*, 2024.